\documentclass[prb,twocolumn,showpacs,floatfix]{revtex4}
\usepackage{graphics,epsfig,amsfonts,amssymb,amsmath}
\begin{document}
\title{Robustness of the magnetoresistance of nanoparticle arrays }
\author{V. Est\'evez}
\author{E. Bascones}
\affiliation{Instituto de Ciencia de Materiales de Madrid,
ICMM-CSIC, Cantoblanco, E-28049 Madrid (Spain).}
\email{vestevez@icmm.csic.es,leni@icmm.csic.es}
\begin{abstract}
Recent work has found that the interplay between spin accumulation and Coulomb
blockade in nanoparticle arrays results in peaky I-V and tunneling
magnetoresistance (TMR) curves and in huge values of the TMR. 
We analyze how these effects are influenced by a 
polarization asymmetry of the electrodes, the dimensionality of the array, the
temperature, resistance or charge disorder and 
long-range interactions. We show that the magnitude and voltage dependence of
the TMR does not change with the dimensionality of the array or
the presence of junction resistance disorder. A different polarization
in the electrodes modifies the peak shape in the I-V and TMR curves but
not their order of magnitude. Increasing the temperature or length of the
interaction reduces to some extent the size of the peaks, being the reduction
due to long-range interactions smaller in longer arrays. Charge disorder
should be avoided to observe large TMR values. 
\end{abstract}
\maketitle
\section{Introduction}
\label{sec:introduction}
A large amount of effort has been devolted to understanding and controlling
the interplay between the electronic current and
magnetism\cite{reviewawschalom,reviewjohnson,reviewfert,spintronics}. Such an
interplay is specially interesting in the presence of charging
effects\cite{cb}. At low temperatures and voltages, the transport through a 
small metallic island is blocked due to the cost in energy to add an electron
to the island. At zero temperature to allow current 
flow a minimum voltage, the threshold voltage $V_{th}$ has to be applied to 
the electrodes. When the island is placed in between two ferromagnetic
electrodes and the spin relaxation time is large, spin accumulation can appear
at the island, i.e. a spin splitting of the chemical potential is created\cite{johnsonprl}. Such
spin accumulation is a non-equilibrium process which appears at finite
voltages induced by the current flow. The spin accumulation happens to
equalize the spin dependent tunneling rates to enter and exit the
nanoparticle. 

When the polarization of source and drain electrodes is the same
spin accumulation appears for antiparallel (AP) arrangement of the electrode
magnetization, but not for parallel (P) orientation and induces a voltage
dependent  TMR\cite{julliere,ono97}, defined as
\begin{equation}
TMR=\frac{R_{AP}-R_P}{R_P}=\frac{I_P-I_{AP}}{I_{AP}}
\end{equation} 
When the spin polarization of the electrodes is different spin accumulation
also appears for P magnetizations and modifies the value of the
magnetoresistance. 
With increasing temperature charging effects are reduced
and the I-V curves and TMR evolve towards the non-interacting values. The 
specific device configuration determines the transport characteristics and
even regions with weak negative differential
conductance  or oscillations in the TMR can be found\cite{barnas98,maekawa98,barnasepl,brataas99,imamura99,yakushiji2005,weymann2005}. 

Charging effects are known to be enhanced in nanoparticle arrays\cite{cb,likharev89,middleton93,nosotrosprb08}. The threshold
voltage increases, being proportional to the number of islands when
interactions are restricted to charges in the same conductor or when there is 
charge disorder\cite{middleton93,nosotrosprb08}. For onsite interactions all the voltage drop happens at the contact
junctions, between islands and electrodes. The large threshold voltage is a
consequence of the charge gradient which has to be created at the inner
junctions, between the nanoparticles, to allow current. Accumulation of charges
is the only way to create a potential drop at these junctions in clean
arrays. Increasing the bias voltage does not influence this potential drop if
the charge state does not change. This changes for long-range
interactions. In this case there are junction dependent voltage 
drops\cite{nosotrosprb08}. On the other hand charge disorder induces voltage 
independent potential steps at the tunnel junctions and modifies the 
threshold voltage. 

Recently we showed that when a nanoparticle array is placed between two
ferromagnetic electrodes, the interplay between charging effects and spin
accumulation has a dramatic effect in the current and 
TMR\cite{nosotroshuge10}. Non-homogeneous 
spin accumulation appears for P orientation of the electrodes. The
inhomogeneity of the spin potential favors a reduction of the threshold
voltage thanks to the spin-dependent potential drops created at the inner
junctions. Because of the relation between spin accumulation and current,
large oscillations appear in the I-V curves.
The spin accumulation decreases when a new conduction channel is open\cite{barnasepl}
producing a reduction in the current  with increasing voltage, and
oscillations appear\cite{nosotroshuge10}. The anomalous
behavior is found below the non-magnetic threshold voltage. Spin
accumulation is also present for AP arrangement of the
magnetizations, but it is fairly homogeneous, so the threshold voltage
and the
overall I-V curves do not change much when compared with the non-magnetic
results. The different threshold voltage for P and AP
configurations results in huge values of the TMR, which are strongly
voltage-dependent. A different behavior in P and AP configurations is also
observed in the relative polarization of the current
$(I_\uparrow-I_\downarrow)/I$. When the 
polarization of source and drain electrodes is equal, the current is not spin
polarized for AP arrangement as it also happens in the single particle case. On
the contrary for P configuration the spin polarization of the current presents
clear oscillations which correlate with the ones found in the current\cite{nosotroshuge10}.

The fact that these huge values of the TMR are associated with 
the (in absence of spin-accumulation) vanishing
differences in energies for tunneling through the inner junctions suggests that
this effect could be extremely sensitive to modifications of the array
characteristics, such as length of the interaction, presence of disorder,
asymmetry in the electrodes polarization, among others or to external 
parameters such as
temperature. In this paper we analyze such sensitivity. We have found that 
the magnitude and voltage dependence of
the TMR do not change with the dimensionality of the array or
the presence of junction resistance disorder. Polarization asymmetry
modifies the peak shape in the I-V and TMR curves but
not their order of magnitude. The size of the peaks is reduced to some extent
with increasing the temperature or length of the
interaction but the effect of long-range interactions is reduced in long 
arrays. On the other hand charge disorder is harmful and should be avoided in
order to observe these effects. 

\section{The system}
\label{sec:model}
We consider an array of metallic nanoparticles placed in between two
ferromagnetic electrodes. Here the word nanoparticle, also called island,  refers to 
confined metallic regions separated from each
other and from the electrodes by tunnel junctions, not necessarily sphere like
metallic particles. They can be nanostructures created by other methods such
as lithography or epitaxy, for example. 

\begin{figure}
\leavevmode
\includegraphics[clip,width=0.5\textwidth]{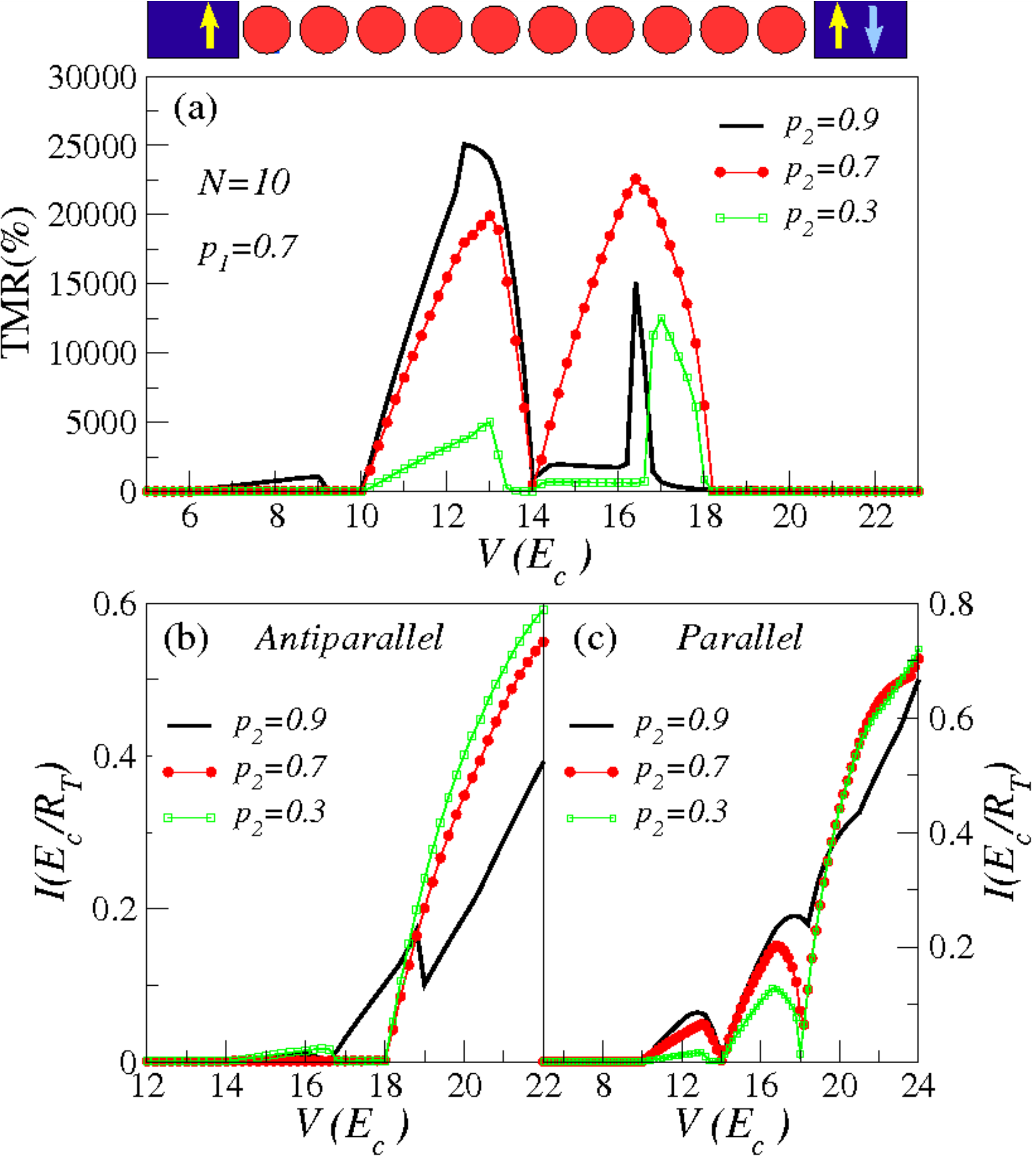}
\caption{Top: sketch of the nanoparticle array under consideration. The arrow
  at the drain refers to the P (light)
  and AP (dark) configurations. (a)TMR as a function of the voltages corresponding to an $N=10$ clean
  array with source polarization $p_1=0.7$ and different drain polarizations
  $p_2$. (b) and (c) I-V curves for the same arrays in (a) for antiparallel and parallel orientation respectively.}
\label{fig:polarizacion1}
\end{figure}

Due to confinement to add an electron to the system costs a charging
energy $E_c$. We restrict ourselves to the classical Coulomb blockade regime 
with $\delta \ll K_BT \ll E_c$ with $\delta$ the level spacing, 
$T$ the temperature and $K_B$ the Boltzmann constant. The arrays
are one-dimensional (1D) and contain $N$ nanoparticles, see sketch
at the top of Fig.~\ref{fig:polarizacion1}, except 
in section \ref{sec:dimensionality} where the effect of the dimensionality is
studied. In this section two-dimensional (2D) arrays with $M
\times N$ particles have $M$ rows  containing each of them $N$ nanoparticles,
see sketch in Fig.~\ref{fig:dosdimensiones}. 

The tunnel resistances which separate the nanoparticles from other
nanoparticles  are  $R_\sigma=2 R_T$ for 
both spin orientations through all the
paper, except in section~\ref{sec:resdisorder} where they are allowed
to vary randomly between two values to  analyze the effect of resistance
disorder in the TMR.  
In equilibrium the nanoparticles are non magnetic but the source and drain
electrodes are ferromagnetic with spin polarization $p_1$ and $p_2$. 
The polarization of the electrodes enters via the tunneling resistance which
depends on spin. At the
contact junctions between the array and the electrodes $R_\sigma=2R_T(1\pm
p_i)^{-1} $ with $\sigma$ the spin index.
Plus (minor) signs are assigned to  majority (minority) spin carriers and 
$i=1,2$ to source and drain
electrodes. Here $2R_T$ is to be substituted by the assigned resistance value
if disorder is included. We consider $p_1=p_2=p$, except in 
section \ref{sec:asymmetry} where the effect of the polarization asymmetry of
the electrodes is studied. 
Transport is treated at the sequential tunneling level with tunneling rates\cite{noteseq}
\begin{equation}
\Gamma_\sigma(\Delta E_\sigma)=\frac{1}{R_\sigma}
\frac{\Delta E_\sigma}{exp( \Delta E_\sigma/K_BT) -1}
\label{eq:sequential}
\end{equation} 
where the electronic charge has been taken equal to unity. $\Delta E_\sigma$
is the change in energy of electrons with spin $\sigma$ due to the tunneling
process and can be written as
\begin{equation}
\Delta E^{\alpha \beta}_\sigma=E^{e-h}_{\alpha \beta}+ (\phi_\beta -\phi_\alpha)
\label{eq:energy}
\end{equation}
where indices $\alpha$ and $\beta$ label the two conductors involved in the
tunneling. $\phi_\alpha$ and $\phi_\beta$ refer to the potential at the
conductor from which the electron leaves and at which the electron arrives,
respectively. $E^{e-h}_{\alpha,\beta}$ is the cost in energy to transfer an
electron between $\alpha$ and $\beta$ assuming that the array is clean, at zero
potential, uncharged and  in equilibrium and equals 
\begin{equation}
E^{e-h}_{\alpha \beta}=E_c^\alpha+E_c^\beta-C^{-1}_{\alpha,\beta}
\label{eq:excitonic}
\end{equation}
For the islands $E_c^\alpha=E_c$, the unit of energy, while at the electrodes
$E_c^\alpha \ll E_C$ and is neglected in the discussion. $C^{-1}$ is the
matrix which describes the interaction between electrons. The diagonal
elements $C^{-1}_{\alpha \alpha}=2 E_c ^{\alpha}$. In Eq. (\ref{eq:excitonic}) 
$C^{-1}_{\alpha \beta}$ measures the attraction of the electron and hole 
created in the conductors due to the tunneling process and is finite only 
when the range of the interaction does not vanish, long-range limit.  Below we concentrate in
the short-range interaction limit $C^{-1}_{\alpha \beta}=\delta_{\alpha\beta}$
in which interactions between electrons are restricted to those charges placed
in the same conductor. In sec~\ref{sec:longrange} we use $C^{-1}_{\alpha
  \beta}=C^{-1}_{\alpha \alpha}e^{|\alpha-\beta|/a_0}$ to analyze how
long-range interactions influence the TMR. An exponential decay is a good
approximation when the islands are capacitively coupled to the neighboring
ones or for screened interactions. Similar approximations have been 
frequently used
in studies of nanoparticle arrays\cite{cb,middleton93,melsen97}. 
Here $a_0$ can be understood as an
effective interaction or screening length and is measured in units of the
nanoparticles center-to-center distance.

At the source and drain electrode the potential are $\phi_{source}=V/2$ and
$\phi_{drain}=-V/2$ respectively. The potential at the islands can be
decomposed in different contributions
\begin{equation}
\phi_\alpha=\tilde\phi^\sigma_{\alpha}+
\phi_{\alpha}^{ch}+\phi_\alpha^{pol}+\phi_\alpha^{dis}
\label{eq:potencial}
\end{equation}
Here $\tilde\phi^\sigma_{\alpha}=N_{\alpha,\sigma}\delta$ with
$N_{\alpha,\sigma}$ the number of excess electrons with spin $\sigma$ at
island $\alpha$. Spin accumulation at the islands is defined as 
$\tilde\phi^\sigma_{\alpha}-\tilde\phi^{-\sigma}_{\alpha}=(N_{\alpha,\sigma}-N_{\alpha,-\sigma})\delta$ and it can be finite under a
non-equilibrium current. 
The charge contribution to the potential depends on the total
excess charges $Q_\alpha=N_{\alpha,\sigma}+N_{\alpha,-\sigma}$ as 
\begin{equation}
\phi_{\alpha}^{ch}=\sum_{\beta=1}^{N}Q_\beta\tilde C^{-1}_{\alpha,\beta} 
\end{equation}
$\tilde C^{-1}_{\alpha \beta}$ is essentially $C^{-1}_{\alpha,\beta}$ plus  a
small modification due to the proximity of the electrodes at a fixed
potential\cite{nosotrosprb08}. 
In the short-range limit $\phi_\alpha^{ch}=2 E_c Q_\alpha$.
$\phi_\alpha^{pol}$ is the polarization potential at the island, induced by
the electrodes at finite bias. In the short-range case it vanishes. This means
that there is no polarization potential drop at the junctions situated between
two islands; the polarization potential drop is only finite at the contact
junctions between an island and an electrode. In the presence of long-range
interactions $\phi_\alpha^{pol}$ is proportional to the bias voltage. For
exponentially decaying interactions 
$\phi_\alpha^{pol}=\frac{V}{2}(e^{-\alpha/a_0}-e^{-|N+1-\alpha|/a_0})$. The
polarization potential drop at the inner junction increases with increasing
the length of the interactions\cite{nosotrosprb08}. The last term in 
Eq.~(\ref{eq:potencial}) $\phi_\alpha^{dis}$ 
is a contribution associated to the presence of random
charges in the substrate or array surroundings. It is finite at the islands
in charge disordered arrays (sec.~\ref{sec:chargedisorder}) and zero in clean arrays, even when there is resistance
disorder. If interactions between the charges are short range
($C^{-1}_{\alpha \beta}=\delta_{\alpha,\beta}$), the set of disorder
potentials $\{\phi_\alpha^{dis}\}$, once the screening of the potential due to the
mobile charges is taken into account, is uniformly distributed in the interval
$-E_c <\phi^{dis}<E_c$. In the presence of long-range interactions the
screened disorder is correlated\cite{nosotrosprb08,eltetoprb05}. Here we only
consider the presence of charge disorder $\phi^{dis}$
in the case of short-range interactions and the first situation applies. 

Except otherwise indicated we consider a 1D array with 
short-range interactions, homogeneous resistances, absence of charge disorder
and take $p=0.7$, $k_BT=10^{-4}$ and
$\delta=10^{-5}$ with the energies  given in units of $E_c$. The value of
$\delta$ does not influence the results as soon as $\delta \ll T$. 

The current is calculated by means of a Monte Carlo simulation described
elsewhere\cite{nosotrosprb08,likharev89,nosotroshuge10}. Simulations become
increasingly costly with increasing temperature. When discussing the effect of
temperature we restrict ourselves to small arrays and temperatures smaller
than $0.01 E_c$.  
\begin{figure}
\leavevmode
\includegraphics[clip,width=0.5\textwidth]{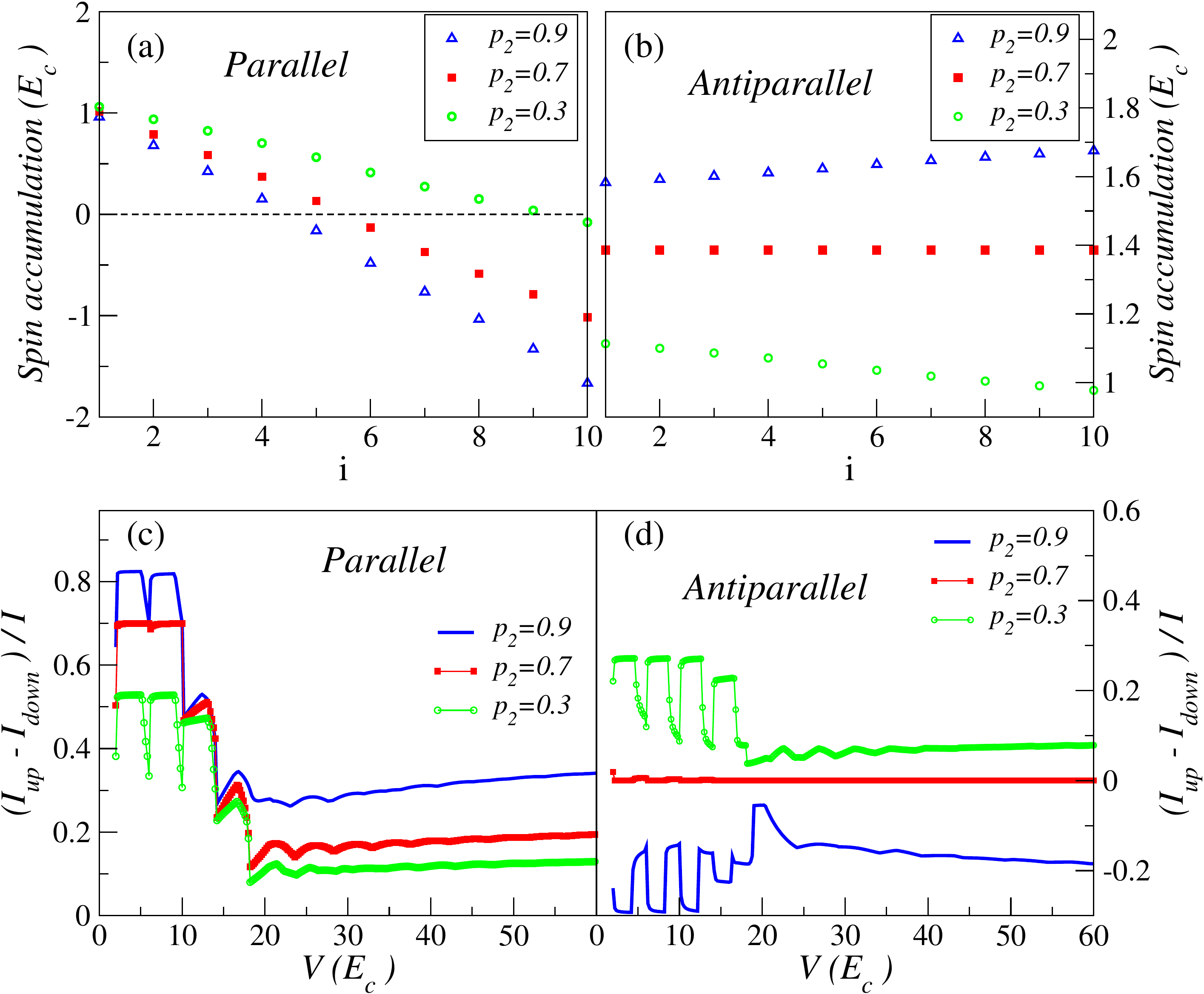}
\caption{(a) and (b) Spin accumulation at the islands as a function of the
  position for the array parameters in Fig.~\ref{fig:polarizacion1} and  $V= 16 E_c^{isl}$
  respectively for parallel and antiparallel arrangement of the electrode
  magnetizations. 
(c) and (d) Spin polarization of the current as a function of the voltage for
the same arrays in (a) and (b).}
\label{fig:polarizacion2}
\end{figure}

\section{Electrode polarization asymmetry}
\label{sec:asymmetry}
In the case of a single island the polarization asymmetry ($p_1\neq p_2$)
induces spin accumulation for P orientation, which is absent if 
$p_1=p_2$. While
the dependence of the TMR on voltage does not change much qualitatively,
 its magnitude depends on the values of $p_1$ and $p_2$,
 saturating at high voltages to $4p_1p_2/(4-(p_1+p_2)^2)$. In  arrays, 
to have  different spin polarization in the electrodes  breaks 
inversion symmetry and is expected to modify the spin potential profile
created along the array. In particular, it can create a spin potential gradient
not only when the electrodes polarizations are parallel, but also when
they are antiparallel. The presence of such a gradient could modify notably
the I-V curves of the AP configuration, allowing current flow below the
metallic threshold and reducing the magnetoresistance in the voltage regime,
where the peaks are observed.

Fig.~\ref{fig:polarizacion1}(a) shows that the peaks in the TMR below the non-magnetic threshold ($V_T=18 E_c$ for $N=10$) do not disappear when spin
asymmetry is present. However, their shape is strongly modified. The values of 
the TMR
have the same order of magnitude as observed when $p_1 =p_2$. Whether they are
larger or smaller depends in a complicated way on the polarizations and on the
voltage. Below $V_T$ and for fixed $p_1$ the current found for parallel
arrangement increases with increasing $p_2$  without altering
much the shape of the I-V curve, see Fig.~\ref{fig:polarizacion1} 
(c). With AP orientation peaks are observed in the
I-V only when $p_1 \neq p_2$, see Fig.~\ref{fig:polarizacion1}(b). Their shape 
is more irregular and 
their height  notably reduced as compared to those found for P-orientation.  
The change in the shape of the TMR curve is mostly due to the change in the
current for AP arrangement.

The differences in the current when compared with the $p_1=p_2$ case originate 
in the different spin gradient created through the array by the spin 
accumulation. The change in spin accumulation between the first and last 
island of the array is much larger
when the electrode magnetizations are P than when they are AP, as 
shown in
Fig.~\ref{fig:polarizacion2} (a) and
(b). For P-orientation when $p_1=p_2$ the sign of the spin accumulation
changes at the center of the array. With spin asymmetry this change of sign
can happen or not and if it does, the change of sign will not happen at the 
array center in a
generic case. For P arrangement and spin asymmetry the sign of the spin
gradient does not depend on the asymmetry, but it is controlled by the
orientation of the electrode magnetization. With AP orientation the spin 
accumulation can increase or decrease
along the array depending on $p_2$ being larger or smaller than $p_1$. At
least for the cases that we have analyzed we have never found a change of sign
in the spin accumulation.      

The polarization asymmetry also modifies  how much spin polarized it is the
current. As it also happens in the single island case, the current in the
AP-arrangement can be spin polarized only if $p_1 \neq p_2$, with the
polarization sign depending on $p_2$ being smaller or larger than $p_1$, see
Fig.~\ref{fig:polarizacion2}(d). For P-orientation it depends on the values of 
the electrode polarizations but does
not change sign, as seen in Fig. 2(c). At high voltages the spin 
polarization saturates to
\begin{equation}
\frac{I^\uparrow-I^\downarrow}{I}=\frac{(p_1\pm
  p_2)(1\mp p_1p_2)}{(1-p_1^2p_2^2)+ N (1-p_1^2)(1-p_2^2)}
\end{equation} 
with upper and lower signs corresponding to P and AP
configurations respectively. 

\section{Dimensionality of the array}
\label{sec:dimensionality}
\begin{figure}
\leavevmode
\includegraphics[clip,width=0.5\textwidth]{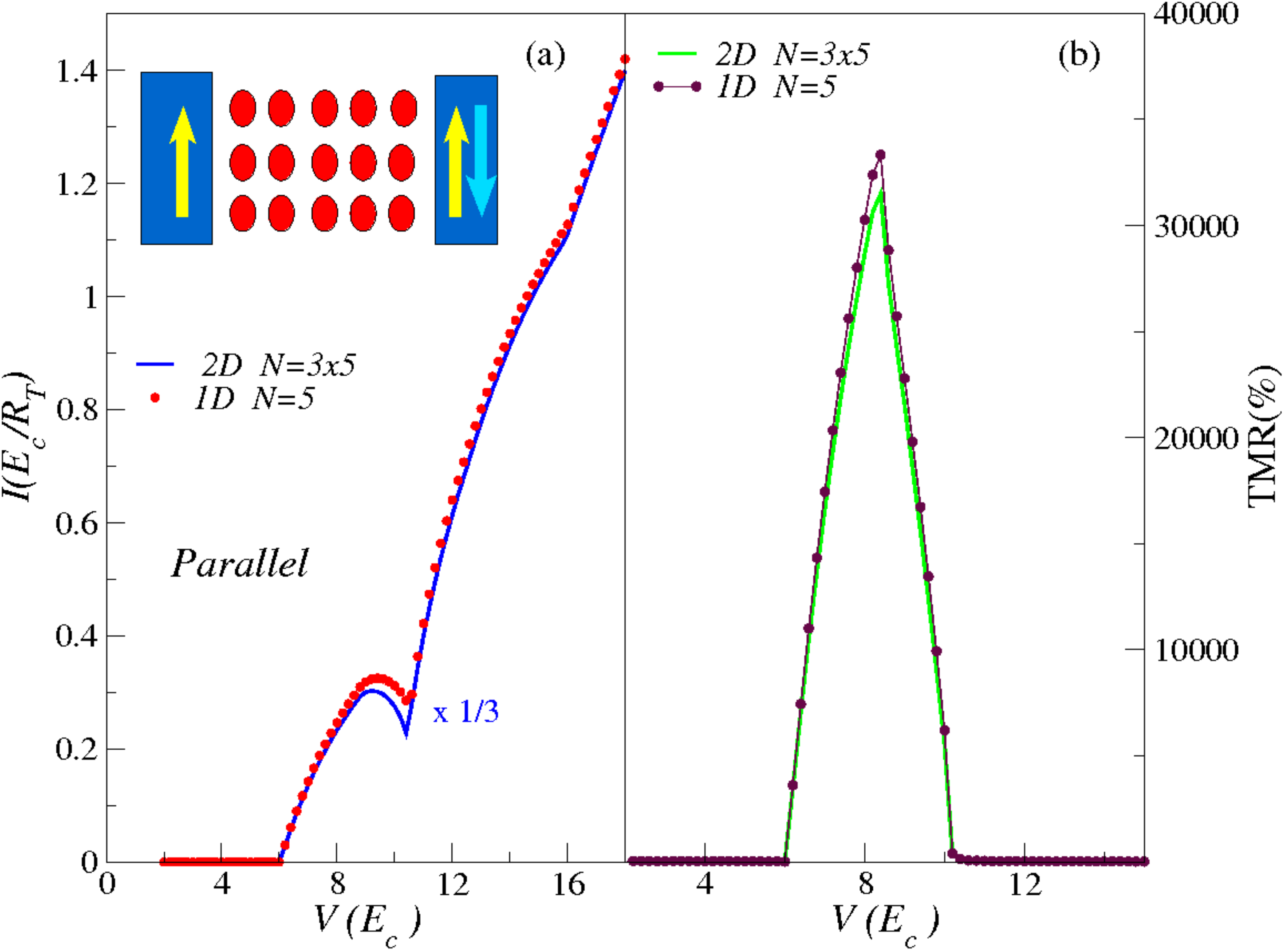}
\caption{(a)I-V curves in one and two dimensional arrays with parallel
  orientation: 1D (N=5), 2D(N=3x5) and p=0.7. The I-V curve for the 2D system has been 
divided by  3, the number of rows.
(b) TMR as a function of the voltage for the arrays in (a). }
\label{fig:dosdimensiones}
\end{figure}
As shown in \cite{nosotroshuge10} to increase the length of a one-dimensional  
array placed between ferromagnetic electrodes from $N=1,2$ to $N=3$ or 
larger has an enormous impact on the I-V curves and on the TMR. Thus it
is sensible to ask what does it happens if we increase the width of the array
to become two-dimensional. For simplicity we assume an square-lattice array, 
see inset in Fig.~\ref{fig:dosdimensiones}. The current through metallic 
two-dimensional 
clean arrays (no charge disorder) is reasonably well described 
by assuming that the different rows conduct in
parallel\cite{nosotros2D10}. If this is true also in the presence of spin 
accumulation no much
change would be expected in the magnetoresistance because the current for both
P and AP would be given by that of a one dimensional array multiplied by the
number of rows. The validity of this approximation is confirmed in Fig. 3
where the TMR and the I-V curve for parallel arrangement for a 1D N=5 array
and for a 2D $3 \times 5$ one (three rows of 5-particles length) are
compared. The agreement
between the I-V curves is very good, once the 2D-array I-V has been
divided by the number of rows. Minor differences are found around the metallic
threshold ($V_T=10$ for $N=5$). As expected the TMR of the 2D arrays follows 
very well the one of the 1D array. A small difference is only seen at the top 
of the peak. The TMR is more sensitive to small changes in the current of the 
AP-configuration (not shown), as this one is much smaller and enters into the 
denominator.

\section{Temperature}
\label{sec:temperature}

\begin{figure}
\leavevmode
\includegraphics[clip,width=0.5\textwidth]{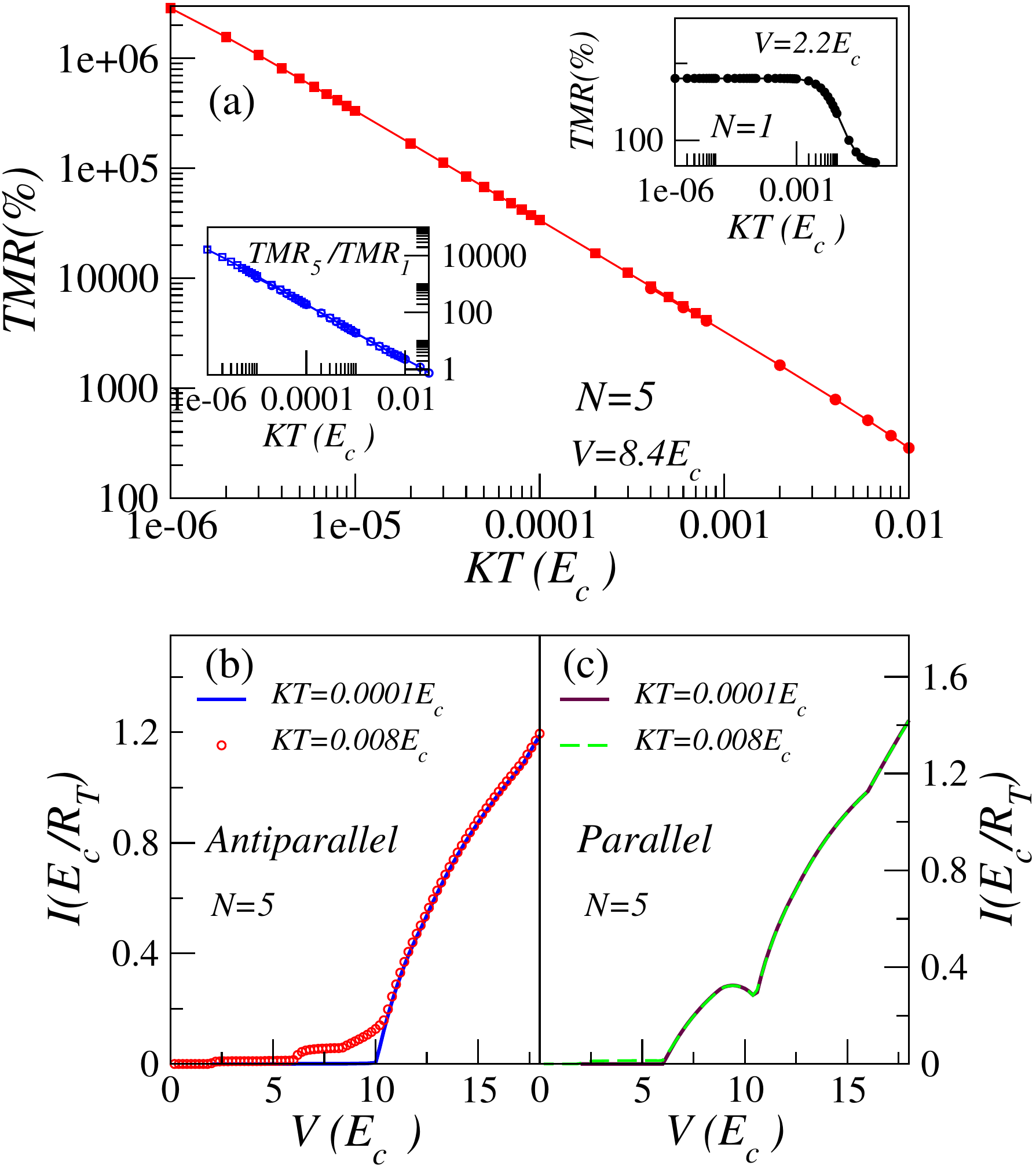}
\caption{(a) TMR as a function of temperature for a clean $N=5$ array and a
  voltage $V=8.4 E_c$. Top inset: TMR as a function of temperature for a
  single island and $V=2 E_c$. It shows a much weaker temperature dependence, 
what is
  emphasized in the lower inset where the dependence of the relative values of
the TMR as a function of temperature are represented. (b) and (c) I-V curves
for an $N=5$ array with AP and P orientations,respectively, for two different 
temperatures. The effect of temperature is much stronger for the AP 
orientation, as expected.}
\label{fig:temperature}
\end{figure}

The high values of the TMR at the peaks are expected to be quite sensitive to
temperature. These values are associated with the small current which flows
through the array below the non-magnetic threshold when the magnetizations are
AP. Such a current extrapolates to zero at zero temperatures. Such vanishing
current at low temperatures is associated with tunneling processes at the
inner junctions with vanishing cost in energy. Increasing the temperature
allows the tunneling event to happen.  
From Eq.~\ref{eq:sequential} the tunneling rate of processes with zero energy cost increases linearly with the temperature. On a first
approximation the current for AP orientation is expected to increases linearly
with the temperature, while the current through an array with P magnetizations
will be much less sensitive because the tunneling processes at the inner
junctions have a finite energy gain provided by the spin potential gradient.
Such dependencies would result in a TMR inversely proportional to the
temperature. Confirmation of this dependence and on the different effect of
the temperature for P and AP orientations is seen in
Fig.~\ref{fig:temperature}. 
      
Note that the strong dependence of the TMR on the temperature is not a
simple consequence of the weakening of Coulomb blockade. As shown in the top 
inset of Fig.~\ref{fig:temperature} (a) the temperature dependence of the TMR corresponding to the single 
island case 
$N=1$ is much weaker at $K_BT\ll E_c$, because the
tunneling processes involved are thermally activated, contrary to the zero
energy cost relevant at the inner junctions of a long array with AP orientation. The different
temperature dependence of the TMR for a single island and a long array is emphasized in the lower inset of the same figure. 

\section{Resistance disorder}
\label{sec:resdisorder}
    The strong exponential dependence of the tunneling probability on the 
  tunneling junction width makes that small differences on the distance
  between the nanoparticles result in large differences between the junction
  resistances. As a consequence resistance disorder is expected to play an
  important role in many devices based on nanoparticle arrays. In this section
  we show that the appearance of peaks in the I-V curves and the large values
  of the TMR are robust against resistance disorder.

\begin{figure}
\leavevmode
\includegraphics[clip,width=0.5\textwidth]{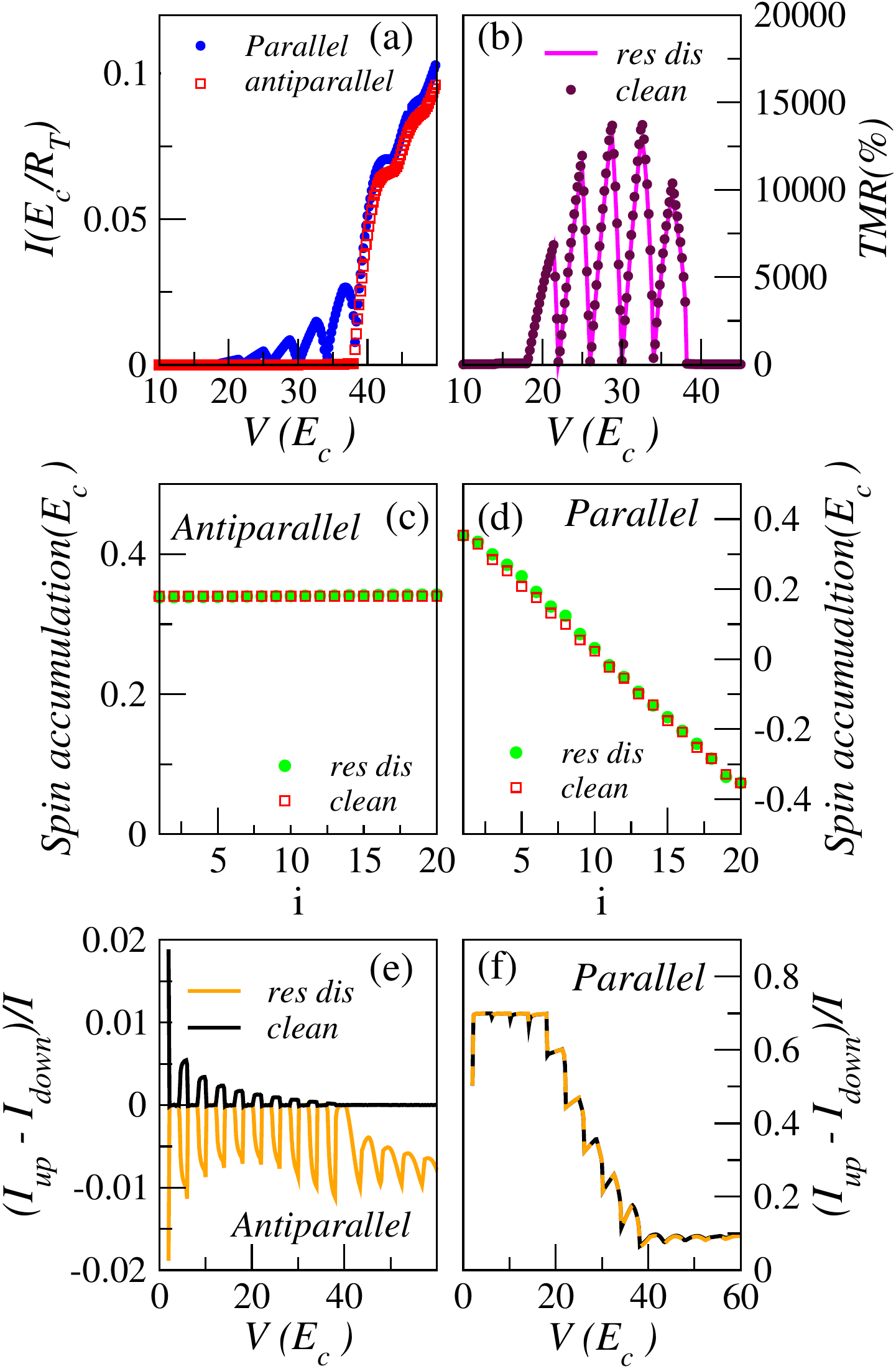}
\caption{(a) I-V curves for  a $N=20$ array  with the junction resistances
  randomly assigned varying between $(10-22)R_T$, for
  both magnetic configurations and $p=0.7$. (b) TMR as a function of the
  voltage for the same array that in (a) compared with the one corresponding
  to a resistance disorder free array.
(c) and (d) Spin accumulation at the islands for $V=38.5 E_c$ with antiparallel 
orientation and $V=20 E_c$ and parallel orientation,
respectively for disorder free and resistance disordered arrays. (e) and (f) Spin polarization of the current for antiparallel
(d) and parallel (e) arrangements for disorder free and
resistance disordered arrays. The curves in all the
graphs correspond to the same $N=20$ arrays with or without resistance
disorder.}
\label{fig:resistancedisorder}
\end{figure}

  Fig.~\ref{fig:resistancedisorder} (a) shows the I-V curves for P and AP
  configurantions corresponding to an $N=20$ 
array  
  with resistance disorder, being the junction resistances
  randomly assigned and varying between $(10-22)R_T$. Due to the different effect of the change in resistances for
  different voltages, these I-V curves are not compared with those found in
  clean arrays. Qualitatively, the main features displayed are the same in 
  both cases. The I-V curve corresponding to AP arrangement shows a threshold
  voltage equal to the one found with non-magnetic electrodes, while the one for
  P-orientation is reduced. Peaks in the current are found at low voltages,
  below $V_T$, the non-magnetic threshold, when the electrode magnetizations are
  parallel. 

  The TMR corresponding to these I-V curves is shown in  
  Fig.~\ref{fig:resistancedisorder} (b) where it is compared with the one
  found in the absence of resistance disorder. In the peaky region of voltages
  no differences are found between the TMR of both systems. 
  The high values of the magnetoresistance are extremely robust against 
  resistance disorder. As could be expected from the shape of the I-V curves
  in Fig.~\ref{fig:resistancedisorder} (a) the spin accumulation, shown in  
  Fig.~\ref{fig:resistancedisorder} (c) and (d) is quite
  homogeneous through an array with AP orientation and decreases from left to
  right with P-orientation. As it happened when all the resistances are equal
  it changes sign at the center of the array, see
  Fig.~\ref{fig:resistancedisorder} (d). For the voltages shown the spin 
  accumulation differs 
  very little when
  compared to the disorder free case.  
  Interestingly, and opposite to what
  happened in uniform arrays, in the AP configuration, at high voltages an 
  spin gradient is formed through the array what produces slight
  differences between the TMR of clean and disordered arrays at these
  voltages, not shown. 

  The spin polarization of the current is shown in 
  Figs.~\ref{fig:resistancedisorder} (e) and (f) for both magnetic
  orientations. In the P-arrangement it equals the one found when there is no
  resistance disorder. The situation changes with AP orientation. 
  At high voltages a finite spin polarization is found in the disordered case, 
  while it vanishes in the disorder-free array. The sign and dependence on 
  voltage
  change for different realizations of the resistance disorder. 
  The finite spin polarization found at
  low voltages is consequence of the finite temperature which allows a small
  current flow below the zero temperature threshold.

\section{Charge disorder}
\label{sec:chargedisorder}  
  Besides resistance disorder, charge disorder can be important in the
  devices. This type of disorder is unavoidable in many systems and specially 
  important in self-assembled nanoparticle arrays. It arises 
  as a consequence of nearby charge impurities
  which create random potentials in the nanoparticles.  These random
  potentials can be described in terms of random background charges spread in
  the interval $(-1/2,1/2)$ in each nanoparticle. 

  Charge disorder has a strong impact on the I-V curves of nanoparticle
  arrays even when the electrodes are non-magnetic. It modifies the threshold 
  voltage, which now depends on the
  particular disorder configuration. In the limit of short range interactions
  discussed here, the threshold voltage, on average, is reduced to the half of
  the value found in a clean-array. Only half of the junctions (upward steps
  in potential) prevent the flow of charge. The voltages at which steps are
  found in the Coulomb staircase regime are also disorder dependent. 

\begin{figure}
\leavevmode
\includegraphics[clip,width=0.5\textwidth]{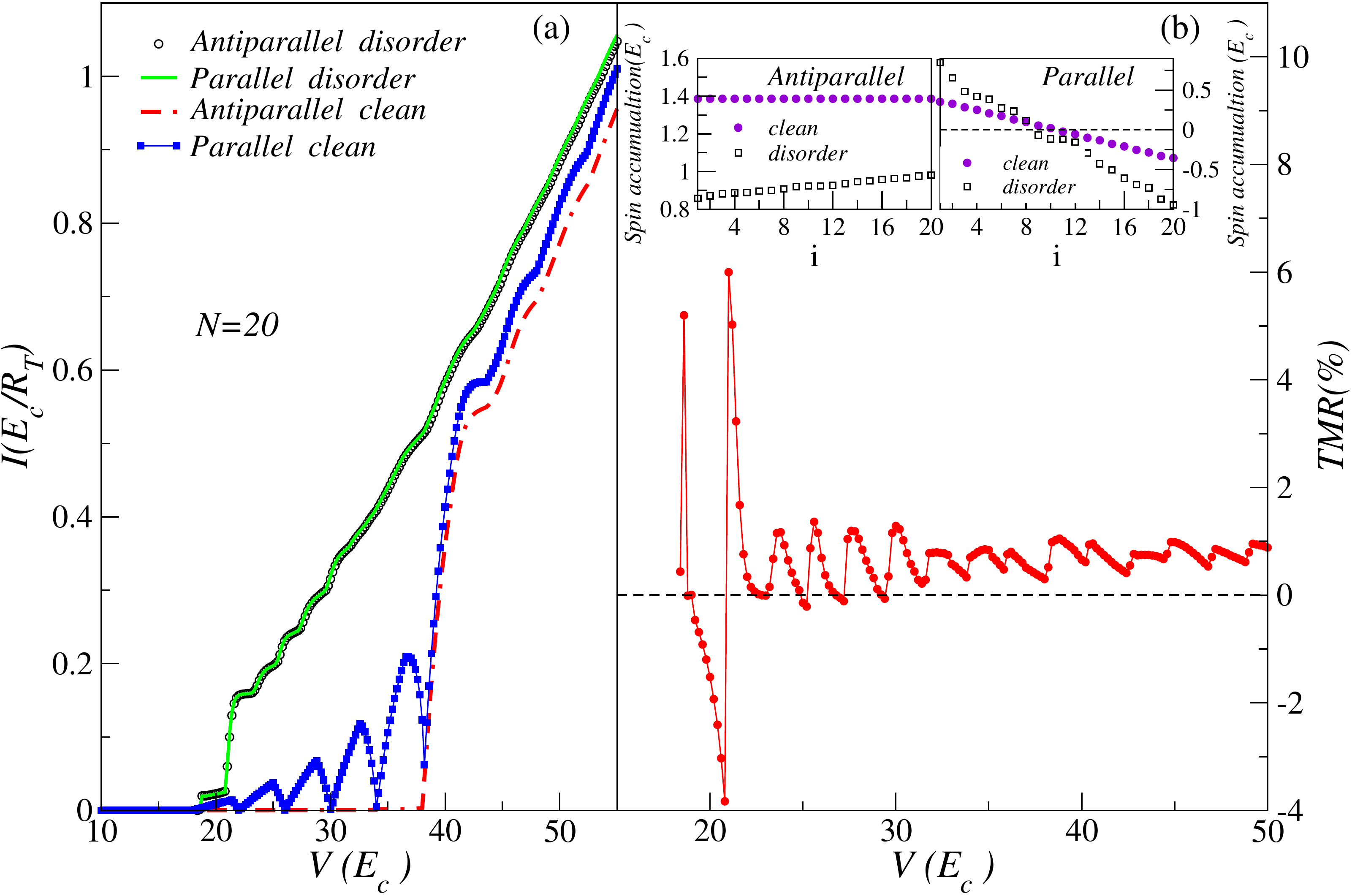}
\caption{(a) I-V curves for $N=20$ clean and charge disordered arrays, with 
source and drain polarization p=0.7 with parallel and antiparallel 
configurations. (b)Main figure:  TMR as a function of the voltage for the 
charge disordered
array used in (a). Insets. Comparison of the spin accumulation gradient
created in the clean and charge disordered arrays in (a) at $V=20 E_c$ 
with antiparallel (left) and parallel (right) orientations.}
\label{fig:chargedisorder}
\end{figure}

  Fig. \ref{fig:chargedisorder} (a) compares the I-V curves of clean and
  disordered arrays when the electrodes are magnetic and their polarizations
  are parallel or antiparallel. The I-V curves of the disordered arrays 
  strongly differ when compared with the clean case. Most significatively, the
  strong dependence of the current on the magnetic orientation of the
  electrodes has disappeared. This is clearly seen in
  Fig. \ref{fig:chargedisorder} (b) which shows a very small
  magnetoresistance. This  figure is to be compared with
  Fig. \ref{fig:resistancedisorder} (b). Random charge disorder has reduced
  the TMR up to four orders of magnitude.

  The large values of TMR found in clean arrays are a consequence of the spin
  potential gradient generated in the array present (absent) when the 
  electrodes 
  magnetizations are parallel (antiparallel). This potential gradient
  allows current to
  flow through the inner junctions. In the absence of such gradient, at zero
  temperature,  the
  current vanishes at low voltages because, while tunneling does not cost 
  energy it also does not gain it.  The suppression of the TMR is not due to a
  reduction of the spin gradient. With charge disorder spin gradients 
  are present, being even larger than in the clean case, see insets in 
  Fig. \ref{fig:chargedisorder} (b). They are finite, but much smaller with
  antiparallel configuration. However, contrary to the clean case, with
  disorder, tunneling through the junctions which prevent current (upward
  steps) costs energy. The change in potential due to spin accumulation is too
  small to overcome such energy cost or to modify significatively the
  tunneling rates through the junctions with downward steps in disorder potential.

\section{Long-range interactions}
\label{sec:longrange}
\begin{figure}
\leavevmode
\includegraphics[clip,width=0.5\textwidth]{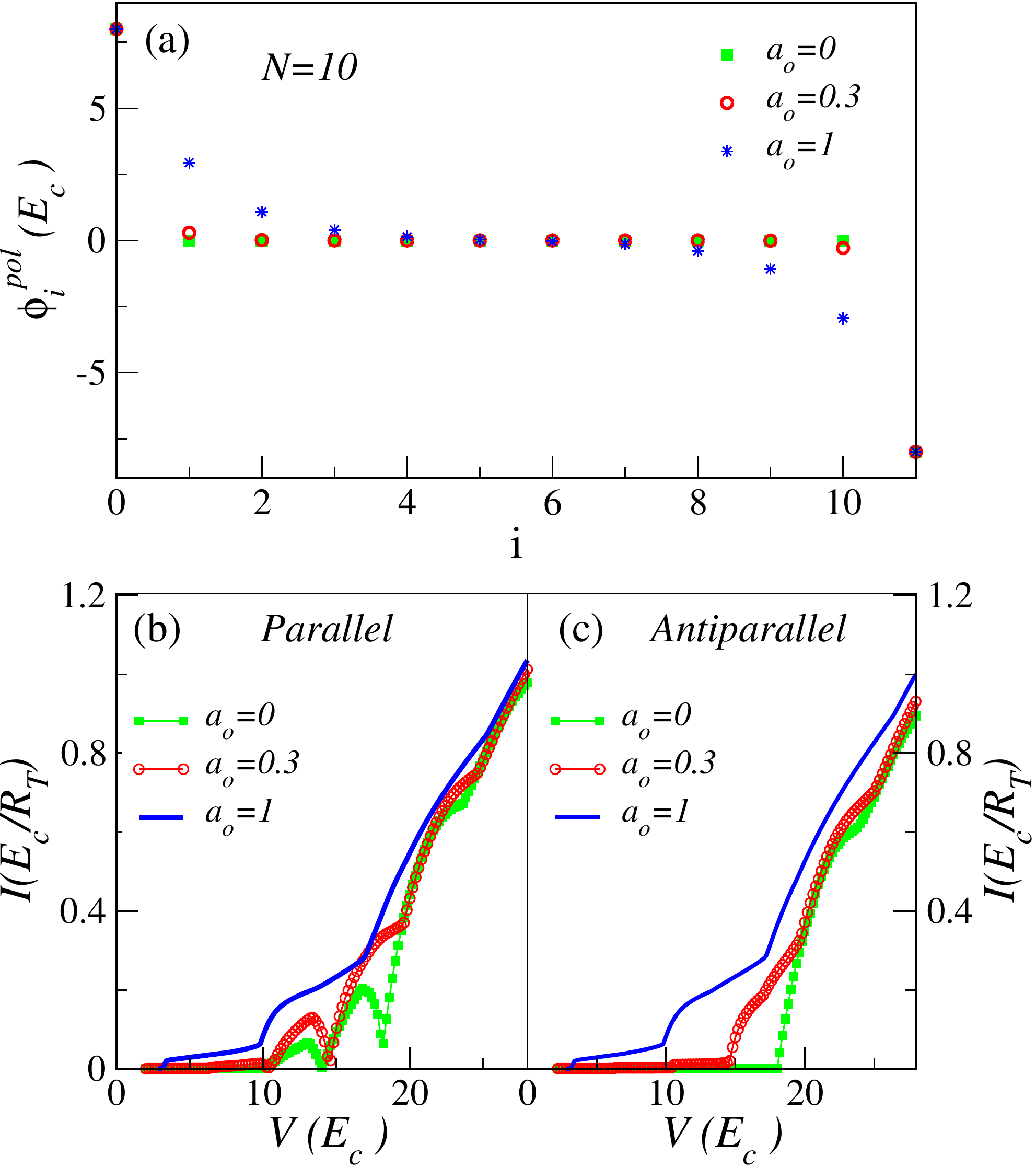}
\caption{(a) Polarization potential at the islands and electrodes for an N=10
  array at $V=16
  E_c$ and several
  values of $a_0$, the range of the interactions, $a_0=0$ refers to the onsite
  interaction limit discussed in previous sections. 
(b) and (c) I-V curves corresponding the same values of $a_0$ and N in (a) for parallel and antiparallel orientation, respectively.}
\label{fig:ivlongrange}
\end{figure}

So far we have studied  onsite interactions in which the interactions
between the electrons are limited to those charges placed in the same
conductor. This case is special because all the voltage drop happens at the
contact junction. There is no polarization potential at the islands. This
makes that at zero temperature and below the threshold corresponding to 
non-magnetic electrodes the current is maintained only by the spin 
accumulation. The potential drop at the inner junctions increases with 
increasing range of the interactions, as can be observed in 
Fig.~\ref{fig:ivlongrange} (a), where an exponentially decaying interaction,
typical of capacitively coupled islands has been used. $a_0$ can be 
understood as a screening length.

For finite range of the interactions and in the absence of charge disorder, 
once a charge has entered the array it is 
able to flow, even in the absence
of charge gradient. The threshold voltage is strongly reduced and given by the
voltage which allows an electron to enter the array or a hole to leave
it\cite{nosotrosprb08,melsen97}.  The role played by the spin accumulation on
the transport is expected to decrease if the interactions are long range. 
The I-V curves corresponding to P and AP orientation, are shown in
Fig.~\ref{fig:ivlongrange} (b) and (c)  for the values of $a_0$ used in (a). The threshold voltage is similar for P and AP
configurations, even if at voltages close to threshold the current is still 
highly suppressed, specially for the smaller values of $a_0$. The potential
drop at the junctions, and correspondingly the I-V curves 
shape are influenced both by the long-range of the interactions which
affects the step width and height and by the spin accumulation responsible 
of the
oscillations.  For small $a_0$
oscillations in the current are observed when the electrode magnetizations are
parallel, but they disappear for large $a_0$. 

\begin{figure}
\leavevmode
\includegraphics[clip,width=0.5\textwidth]{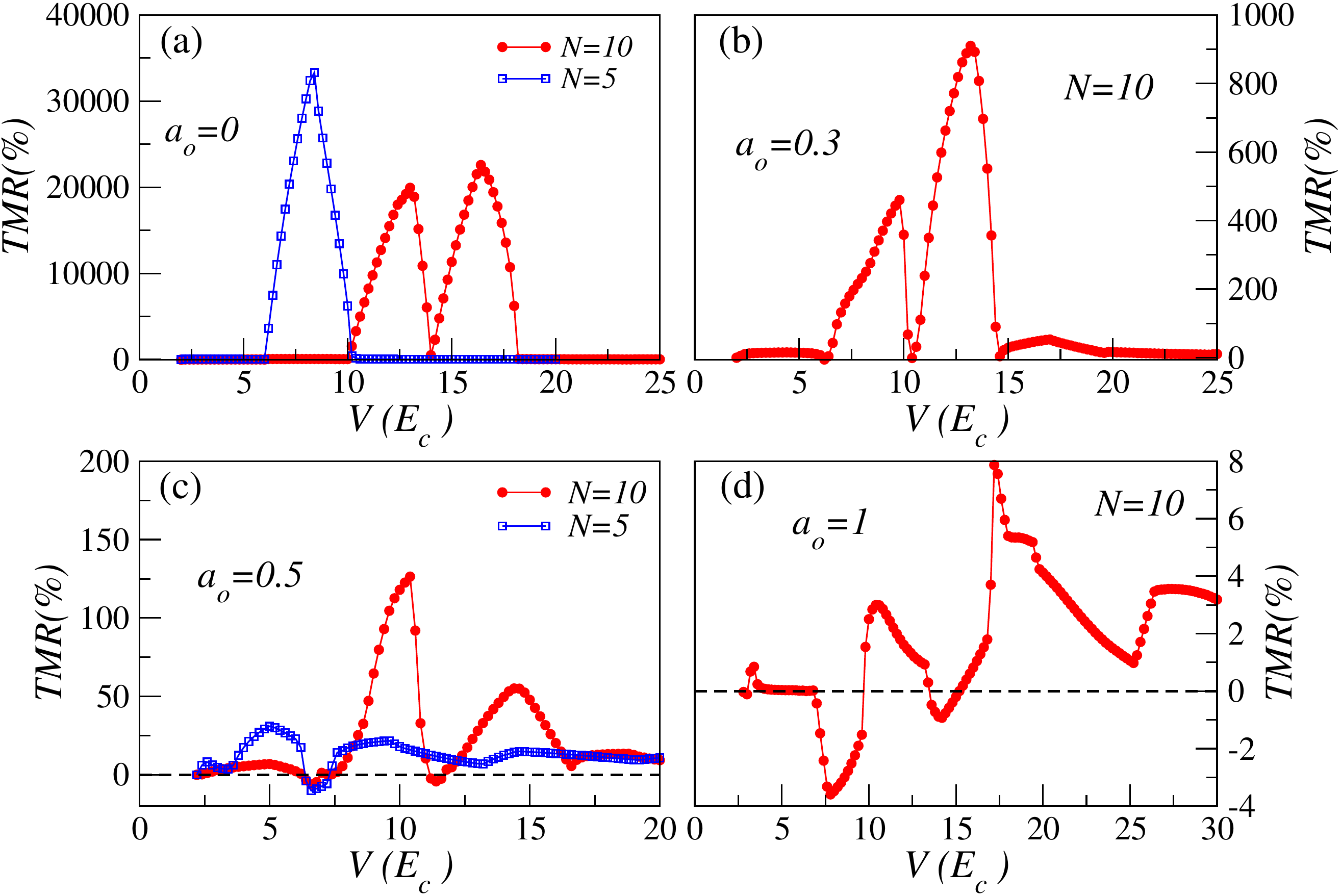}
\caption{(a) to (d) TMR as a function of voltage for an $N=10$ array for
  $a_0=0,0.3,0.5$ and $1$, respectively. In (a) and (c) the TMR of an $N=5$ 
  array with the same $a_0$ is included. The reduction of the
  oscillations amplitude with $a_0$ is faster in shorter arrays.}
\label{fig:tmrlongrange}
\end{figure}

The effect of increasing the range of the interaction on the TMR is plotted in 
Fig.~\ref{fig:tmrlongrange}. As expected 
the amplitude of the oscillations in the TMR is clearly reduced as $a_0$ 
increases.  Note the different scale of the TMR axis in the
plots. For $N=10$ and $a_0=1$ the TMR almost
vanishes. On the other hand, as seen in Fig.~\ref{fig:tmrlongrange}~(c) the reduction of the TMR for a given
interaction length depends strongly on the array length. As longer it is the
array the reduction in the magnitude of the oscillations is weaker. 
This dependence appears because given an interaction length the polarization potential drops at the junctions between 
particles is larger for smaller $N$. 

\section{Summary and Discussion}
\label{sec:summary}
In summary, we have seen that the oscillations in the TMR and in the I-V
curves are robust to most of the  modifications that can appear in an
experiment, when compared to the ideal case studied in
\cite{nosotroshuge10}. In particular if source and drain have different spin
polarizations the magnitude of the oscillations does not change
much. The main effect induced by the polarization symmetry is a change in the
peak shape and the appearance of peaks when the magnetizations are
antiparallel. To go from a 1D to a 2D-square lattice array does not affect the
TMR to a first approximation, while the current is increased following the
number of rows in the 2D array. A similar situation is found in arrays with
resistance disorder. 

The amplitude of the oscillations and the TMR decreases with temperature so,
the temperature should be kept as small as possible in an experiment as compared to the
charging energy. Both large $E_c$ and small values of $K_BT$ are
convenient.   

In many arrays the long-range part of the interaction is screened by mobile
charges in neigbour conductors. If this does not happen the TMR can be
suppressed with respect to the onsite interactions case. This effect is less
important in long arrays. Thus, if the length of the interaction has to
be taken into account the oscillations and large values of the TMR will be
better observed in long arrays.    

The most harmful effect is charge disorder. In some systems like
self-assembled arrays this kind of disorder seems unavoidable because of
charges quenched on the substracted. However, in other devices like 
epitaxially grown pillars the disorder will be probably less important, what
makes them a better candidate to observe these effects.

We thank M.A. Garcia for useful discussions. 
Funding from Ministerio de Ciencia e Innovaci\'on through
Grants No. FIS2008-00124/FIS, FPI fellowship  and
from Consejer\'ia de Educaci\'on de la Comunidad Aut\'onoma de Madrid and CSIC
through Grants No. CCG08-CSIC/ESP3518, PIE-200960I033 is 
acknowledged.


\begin{thebibliography}{99}
\bibitem{noteseq}To use this function a broadening of the energy levels of
  order $\delta/2$ is
  assumed.
\end{thebibliography}
\end{document}